\documentclass[12pt,preprint]{aastex}
\usepackage{amsmath}

\shorttitle{MASSLESS CLASSICAL ELECTRODYNAMICS}%
\shortauthors{Ibison}%

\begin{document}
\title{MASSLESS CLASSICAL ELECTRODYNAMICS}
\affil{Institute for Advanced Studies at Austin}%
\affil{4030 West Braker Lane, Suite 300, Austin, Texas 78759}%
\email{ibison@earthtech.org}%
\email{accepted for publication in Fizika A}%
%
\begin{abstract}
%
In the direct action form of classical EM we give the equation of motion for a classical massless bare charge without self-interaction
in the presence of an external field. That equation permits superluminal speeds and time-reversals, and so is a realization of the
Stueckelberg-Feynman view of electrons and positrons as different segments of a single trajectory. We give a particular solution to a
one body problem, and briefly discuss some aspects of the two-body problem. There is some discussion of the historical context of this
effort, including the direct action and absorber theories, and some speculation on how the massless bare charge may acquire mass, and
how these findings impact the problem of singular self-action.
\end{abstract}


\label{firstpage}

%
\section{Introduction}
%
Stueckelberg \cite{1:stuckelberg 1, 2:stuckelberg 2} and Feynman \cite{3:feynman positrons} suggested that all electrons and positrons
are the same particle undergoing time reversals. Classical electrodynamics (CED) prohibits a particle passing from a sub-luminal to
superluminal speed and undergoing a time-reversal, and therefore apparently cannot implement the Stueckelberg-Feynman conjecture.
However, this prohibition applies only to charged particles possessing an intrinsic rest mass, whereas we show here that CED without
the traditional inertial-mass action permits transitions from sub-luminal to superluminal speeds by the same particle.

We take as a starting point here that the bare charge is free of both self-action and compensating forces. Since radiation reaction is
also an action of the particle's fields upon itself, it follows that the bare charge is not subject to radiation reaction, and, from
energy and momentum conservation, that the secondary radiation emitted by the particle can carry no independent energy or momentum.
These requirements are satisfied by the time-symmetric, direct-action without self-action form of electromagnetism that originated
with Schwarzschild, Tetrode, and Fokker in the early part of the last century \cite{4:schwarzschild, 5:tetrode, 6:fokker}. Those early
presentations lacked an explanation for exclusively retarded radiation and the radiation reaction on the source. Subsequently Dirac
\cite{7:dirac} showed that radiation reaction arises if the advanced fields are set to zero, for which the Wheeler and Feynman
\cite{8:feynman absorber, 9:feynman direct action} absorber theory gave a physical justification. Here however, unlike those works,
the charge sources will not be ascribed an intrinsic (now necessarily non-electromagnetic) inertial mass, because the focus of this
document is on the behavior of classical charges in their allegedly pre-mass condition.

In direct action electrodynamics, EM fields, if they are used at all, are purely mathematical devices for conveying `interaction'
between pairs of charged particles; there are no independent (vacuum) fields. A classical particle then has no self energy - though
there seems to be no simple way to rid the corresponding quantized theories of divergences \cite{13:hoyle 1, 14:hoyle 2, 15:davies
direct action 1, 16:davies direct action 2}. We will need to revisit this issue, however, in Section \ref{sec:Self action}.

Initially we will presume a state of affairs at zero Kelvin wherein the advanced and retarded fields have equal status, wherein one is
not cancelled and the other enhanced at the location of the particle. The viability of the direct action formulation at elevated
(non-zero) temperatures however demands a plausible explanation for the predominance of retarded radiation. The particular
implementation of direct-action EM advocated by Wheeler and Feynman in order to explain the emergence of time-asymmetric retarded
radiation is contingent on the existence of relatively cold distant absorbers of radiation on the future light cone. These absorbers
provide a thermal sink, permitting radiation energy and momentum to flow away from local sources, thereby performing the role
traditionally played by the vacuum degrees of freedom. The flow of radiation energy is properly accompanied by a reaction back upon
the source, performing the role traditionally played by the self-fields. The time-asymmetric boundary condition (cold absorbers on the
future light cone) also succeeds in explaining how time-asymmetric radiation arises in an intrinsically time-symmetric theory. (The
equations of classical EM - with or without field degrees of freedom - are of course intrinsically time-symmetric.) In order to
guarantee complete absorption on the future light cone, the absorber theory places constraints on the rate of cosmological expansion,
which according to Davies, \cite{17:davies absorber, 18:davies book}, are satisfied only by the oscillating Friedman cosmology - which
is now out of favor. Davies claims that the constraints are also satisfied by the steady-state theory which, however, has also fallen
out of favor due to its failure to explain the Cosmic Microwave Background.  In his comprehensive review of the subject, Pegg,
\cite{19:pegg}, contests Davies' claim that radiation in the steady-state cosmology is perfectly absorbed by future matter, citing the
fall-off with increasing wavelength - and therefore red-shift - of absorbtion in a plasma. However, there may still be sufficient
absorbtion in the steady-state cosmology through the action of Thomson scattering, which remains independent of frequency for low
frequencies, and though elastic, removes energy from the primary wave.

Wheeler and Feynman also proposed an alternative to the theory of future absorbers wherein the future is transparent and the history
is perfectly absorbing to advanced radiation emitted from the present \cite{8:feynman absorber}. This possibility appears to be
consistent with the open (and flat) Friedman cosmologies, and the steady-state theory.

The investigation of a massless condition of a classical charge would benefit from some indication of how the observed mass might
arise classically. (The Higgs mechanism, which otherwise satisfies that requirement, is excluded.) In particular we have in mind
attempts to explain inertial mass in electromagnetic terms - an endeavour towards which Feynman was sympathetic \cite{20:feynman
mass}. It is not correct to include in such attempts the now discarded models of the classical electron by Poincar\'{e}
\cite{21:poincare} and updated by Schwinger \cite{22:schwinger} because these are not solely electromagnetic; they rely upon
non-electromagnetic forces to hold the particle together. Of greater relevance are recent efforts to explain inertial mass by calling
upon a special role for the ZPF \cite{23:haisch 1} - \cite{30:ibison ZPF}. An alternative suggestion, favoring a fully
\textit{non-local} electromagnetic origin of, specifically, electron mass, is given in Section \ref{ref:Electron mass}.
%
\section{Equation of motion}\label{sec:Equation of motion}
\subsection{Action and the Euler equation}\label{sec:Action and the Euler equation}
%
In the following we begin with the simplest possible version of classical electromagnetism: with the particle stripped of intrinsic
mass, and the fields stripped of vacuum degrees of freedom radiation reaction, and time-asymmetry. We use the convention
$u^{a}v_{a}=u_{0}v_{0}-{\bf u\cdot v}$ and Heaviside-Lorentz units with $c=1$ except for the explicit calculation in Section
\ref{sec:Motion near a charge with magnetic dipole moment}.

The contribution to the action from a single massless charge, assuming the fields are given, is just
\begin{equation} \label{interaction}
I=-\int d^{4}yA_{\mu}\left(  y\right)  j^{\mu}\left(  y\right) =-e\int d\lambda A_{\mu}\left(  x\left(  \lambda\right) \right)
u^{\mu}\left(\lambda\right)
\end{equation}
where we have used that the 4-current due to a single charge is
\begin{equation} \label{4-current}
j^{\mu}\left(  y\right)  =e\int d\lambda u^{\mu}\left(  \lambda\right) \delta^{4}\left(  y-x\left(  \lambda\right)  \right)  ;\quad
u^{\mu}\left( \lambda\right)  \equiv\frac{dx^{\mu}\left(  \lambda\right)  }{d\lambda}
\end{equation}
where $\lambda$ is any ordinal parameterization of the trajectory, and where $x$ and $y$ are 4-vectors. From its definition, the
current is divergenceless for as long as the trajectory has no visible end-points. In the event that it is necessary to refer to other
particles, let the particular source that is the subject of Eq.~(\ref{interaction}) have label $l$. And consistent with the maxim of
direct action without self action, the potential in Eq.~(\ref{interaction}) must be that of other sources, which therefore can be
written
\begin{equation} \label{total potential}
A_{\overline{l}}^{\mu}= \sum_{k\,;\,\,k\neq l} A_{k}^{\mu}\quad;\quad A_{k}^{\mu}=G\ast j_{k}^{\mu}
\end{equation}
where the bar over the $l$ signifies that the potential is formed from contributions from all sources except the $l^{th}$ source;
where $G$ is the time-symmetric Green's function for the wave equation; and the $\ast$
represents convolution%
\begin{equation} \label{single potential}
A_{k}^{\mu}\left(y\right)=\frac{1}{4\pi}\int d^{4}x\delta\left(\left(y-x\right)^{2}\right) j_{k}^{\mu }\left( x\right) =
\frac{e_{k}}{4\pi}\int d\kappa \delta\left(  \left(  y-x_{k}\left( \kappa\right) \right) ^{2}\right) u^{\mu}_{k}\left(\kappa\right)
\,,
\end{equation}
where $\left(  y-x\right)  ^{2}\equiv\left(  y-x\right)  ^{\mu}\left( y-x\right)  _{\mu}$. These `particle-specific' fields are the
same as those by Leiter \cite{31:leiter}. Putting this into Eq.~(\ref{total potential}) and putting that into Eq.~(\ref{interaction})
and then summing over $l$ would cause each unique product (i.e. pair) of currents to appear exactly twice. Considering every particle
position as an independent degree of freedom, the resulting total action, consistent with the action Eq.~(\ref{interaction}) for just
one degree of freedom, is therefore
\begin{equation} \label{total interaction}
I_{all} = -\sum_{k,\,l;\,k\neq l}\frac{e_{k}e_{l}}{8\pi}
 \int d\kappa\int d\lambda\delta\left(  \left(
x_{k}\left(  \kappa\right)  -x_{l}\left( \lambda\right)  \right)  ^{2}\right)  u_{k}^{\mu}\left( \kappa\right) u_{\mu
l}\left(\lambda\right)\,.
\end{equation}
Equivalent to the supposition that the potential from the other sources, $k\neq l$, is given, is that the fields are in no way
correlated with the motion of the single source responsible for the current $j$ in Eq.~(\ref{interaction}). Therefore this
investigation may be regarded as an analysis of the state of affairs pertaining to the first of an infinite sequence of iterations of
the interaction between the $l^{th}$ current and the distant $k\neq l$ currents responsible for the fields.

With the fields given, the Euler equation for the (massless) lone particle degree of freedom in Eq.~(\ref{interaction}) is simply that
the Lorentz force on the particle in question must vanish:%
\begin{equation} \label{lorentz force}
F_{\overline{l}}^{\nu\mu}u_{\mu l}=0
\end{equation}
where $F$ is the EM field-strength tensor, wherein the fields $\mathbf{E}$ and
$\mathbf{B}$ are to be evaluated along the trajectory. In 3+1 form, and
omitting the particle labels, this is%
\begin{equation} \label{lorentz force in 3+1}
\frac{dt\left(\lambda\right)}{d\lambda} \mathbf{E} \left(\mathbf{x}\left(\lambda\right),t\left(\lambda\right)\right)
+\frac{d{{\mathbf{x}}\left(  \lambda\right)  }}{d{\lambda}} \times{
\mathbf{B}\left(\mathbf{x}\left(\lambda\right),t\left(\lambda\right)\right)={\mathbf{0}}}
\end{equation}
where $\mathbf{E}$ and $\mathbf{B}$ can be found from the usual relations to
$A$.
%
\subsection{Confinement to a nodal surface}\label{sec:Confinement to a nodal surface}
%
For Eq.~(\ref{lorentz force}) to have a solution, the determinant of $F$ must vanish, which gives
\begin{equation} \label{E.B=0}
S\left(x\left(\lambda\right)\right) \equiv \mathbf{E} \left(x\left(\lambda\right)\right) \cdot{\mathbf{B}
\left(x\left(\lambda\right)\right)}=0 \,.
\end{equation}
Note that though this condition imposes a constraint on the values of the fields, it does so only on the trajectory itself.
Consequently, Eq.~(\ref{E.B=0}) permits the interpretation that it is a constraint on the set of possible paths that a trajectory can
take through a given, and perfectly general, set of fields. It is consistent with the condition that the Lorentz force on the particle
must vanish - recognized as the constraint on the fields such that there exist a frame in which the electric field is zero. In an
environment of arbitrary field variation, Eq.~(\ref{E.B=0}) selects the surface upon which a charge source may conceivably see no
electric field in its own frame. Accordingly, Eq.~(\ref{E.B=0}) will be regarded as a constraint on the initial conditions - the
location of the charge source at some historical time. Only if the source is initially placed upon this surface, can a trajectory
exist consistent with the presumption of masslessness and consequent vanishing of the Lorentz force. At present it is not known if
Eq.~(\ref{E.B=0}) permits multiple, unconnected, closed surfaces, connoting localization of the particle.

Since Eq.~(\ref{E.B=0}) is required to be true for all $\lambda$-time along the trajectory, it must be true that all the derivatives
with respect to $\lambda$ of the function $S$ are zero. In particular, in order to solve Eq.~(\ref{lorentz force}), we will need that
\begin{equation} \label{dphi is zero}
\frac{dS}{d{\lambda}}=u^{\mu}\partial_{\mu}S=0 \,,
\end{equation}
which is just the condition for particle to remain on the surface.
%
\subsection{The trajectory of a massless charge}\label{sec:The trajectory of a massless charge}
%
Writing Eq.~(\ref{lorentz force in 3+1}) in the form
\begin{equation}
\dot{t}{\mathbf{E}}+{\mathbf{\dot{x}}}\times{\mathbf{B}}={\mathbf{0}}
\label{LF}
\end{equation}
where dots indicate differentiation with respect to $\lambda$, it may be observed that ${{\bf \dot{x}}\cdot{\bf E}=0}$; the velocity
is always perpendicular to the local electric field. The vectors ${\bf E}$ and ${\bf B}$ are mutually orthogonal and both of them are
orthogonal to ${\bf E}$ because ${{\bf E\cdot B}=0}$. Therefore they can serve as an orthogonal basis for the velocity:
\begin{equation}
{\bf \dot{x}}=\alpha{\bf E}\times {\bf B} +\beta{\bf B}
\label{xdot}
\end{equation}
where $\alpha$ and $\beta$ are undetermined coefficients. Substitution of this expression into Eq.~(\ref{dphi is zero}) gives
\begin{equation} \label{alpha}
\begin{aligned}
\dot{t}{\mathbf{E}}+\alpha\left(  {{\mathbf{E}}\times{\mathbf{B} }}\right)  \times{\mathbf{B}}
    &= \dot{t}{\mathbf{E}}+\alpha\left(
\left( {{\mathbf{E{\cdot}B}}}\right)  {\bf B}-B^{2}{\bf E}\right)={\bf 0} \\
  \Rightarrow \alpha  &=\dot{t}/B^{2}
\end{aligned}
\end{equation}
unless ${\bf B}$ is zero. Substitution of Eqs.~(\ref{xdot}) and (\ref{alpha}) into Eq.~(\ref{dphi is zero}) then gives
\begin{equation} \label{beta}
\begin{aligned}
\dot{t}\frac{\partial S}{\partial t}+\frac{\dot{t}}{B^2}\left( {\bf E}\times{\bf B}\right)  \cdot \nabla S+\beta{\bf B}\cdot \nabla
S=0 \\
\Rightarrow\beta=-\frac{\dot{t}}{B^{2}}{\bf B}\cdot \nabla S \left( \left({\bf E}\times {\bf B}\right) \cdot \nabla S+B^{2}
\frac{\partial S}{\partial t}\right)
\end{aligned}
\end{equation}
unless ${{\bf B\cdot}}\nabla S$ is zero. With Eqs.~(\ref{alpha}) and (\ref{beta}), the velocity, Eq.~(\ref{xdot}), is
\begin{equation} \label{xdot-2}
\begin{split}
{\bf \dot{x}}
    &= \frac{\dot{t}}{B^{2}{\bf B\cdot}\nabla S} \left( \left( {\bf B}\cdot \nabla S\right){\bf E} \times{\bf B}-\left(
        \left(  {\bf E}\times {\bf B}\right) \cdot \nabla S+B^{2}\frac{\partial S}{\partial t}\right) {\bf B}\right) \\
    &=\frac{\dot{t}}{B^2{\bf B}\cdot\nabla S}\left(  \left( {\bf B}\times \left({\bf E}\times {\bf B}\right) \right) \times \nabla
        S-B^{2}{\bf B}\frac{\partial S}{\partial t}\right) \\
    &=\frac{\dot{t}}{B^{2}{{\mathbf{B\cdot}}}\nabla S}\left(  \left(
        B^{2}\mathbf{E-}\left(  {{\mathbf{E{\cdot}B}}}\right)  \mathbf{B}\right)
        {\times}\nabla S-B^{2}{\bf B}\frac{\partial S}{\partial t}\right)\\
    &=\frac{\dot{t}} {{\bf B\cdot}\nabla S}\left(  {\bf E}\times \nabla S-{\bf B}\frac{\partial S}{\partial t}\right) \,.
\end{split}
\end{equation}
With this the 4-velocity is%
\begin{equation} \label{w}
u^{\mu}=f\left(  x{\left(  \lambda\right)}\right)  w^{\mu}
\end{equation}
where, using the convention that a 4-vector with a non-repeated symbolic-non-numerical index, e.g. $u^{\mu}$, means the set of 4
coordinates rather than a single element, and
\begin{equation} \label{w defn}
w^{\mu}\equiv-\widetilde{F}^{\mu\nu}\partial_{\nu}S=\left( {{\mathbf{B\cdot}}}\nabla
S,{{\mathbf{E}}\times\nabla}S-{{\mathbf{B}}}\frac{\partial S}{\partial t}\right)
\end{equation}
where $\widetilde{F}$ is the dual of $F$, i.e. $\widetilde {F}^{ab}=\epsilon^{abcd}F_{cd}$ where $\epsilon$ is the totally
anti-symmetric tensor \cite{32:jackson book}, and where
\begin{equation}
f\left(x{\left(\lambda\right)}\right)=\dot{t}/{\bf B}\cdot \nabla S \label{f(x,lambda)}
\end{equation}
is an arbitrary function, undetermined by Eq.~(\ref{LF}). That
\begin{equation}
u^{\mu}=-f \widetilde{F}_{\mu\kappa}\partial^{\kappa}S
\end{equation}
solves Eq.~(\ref{lorentz force}) is easily confirmed upon substitution, whereupon
\begin{equation} \label{Fu}
F^{\nu\mu}u_{\mu}=-fF^{\nu\mu} \widetilde{F}_{\mu\kappa}\partial^{\kappa}S\,.
\end{equation}
But it is easily computed that
\begin{equation}
F^{\nu\mu}\widetilde{F}_{\mu\kappa}=\delta_{\kappa}^{\nu}S \,,
\end{equation}
so Eq.~(\ref{Fu}) is
\begin{equation}
F^{\nu\mu}u_{\mu}=-fS\partial^{\nu}S
\end{equation}
which is zero on $S=0$, as required.
%
\subsection{Segmentation into a sequence of 4-vectors}\label{sec:Segmentation into a sequence of 4-vectors}
%
One obtains from Eq.~(\ref{xdot-2}) that
\begin{equation}
{\bf v}\left(  {{\mathbf{x}},t}\right)  =\frac{d {\bf x/} d\mathbf{\lambda}}{dt/d\mathbf{\lambda}}=\frac{{\bf E} \times\nabla S-{\bf
B}\partial S\mathbf{/}\partial t} {{\bf B}\cdot \nabla S} \label{v ito E and B}
\end{equation}
is the ordinary velocity of the trajectory passing through $\left(t{\left(\lambda\right),{\mathbf{x}}\left(\lambda\right) }\right)$.
The right hand side is an arbitrary function of $\mathbf{x}$ and $t$, decided by the fields. In general, Eq.~(\ref{v ito E and B})
will not admit a solution of the form ${\bf x=f}(t)$ since the solution trajectory may be non-monotonic in time. With this caveat in
principle Eq.~(\ref{v ito E and B}) may be solved to give the trajectory, and is therefore a complete description for a single
trajectory as it stands, provided one ignores the {\it sense} (see below).

Let us suppose for now that the trajectory is sparse, so that $u^{\mu}$ defined in Eq.~(\ref{w}) cannot be a 4-vector field, because
it is not defined off the trajectory. Then one would like to parameterize the trajectory in a Lorentz invariant way, so that $u$ along
the trajectory is a (Lorentz) 4-vector. This requires that the norm
\begin{equation} \label{u-squared}
u^{\mu}u_{\mu}=f^{2}\left(  x{\left(  \lambda\right) }\right) w^{\mu}\left(  x{\left(  \lambda\right)  }\right) w_{\mu}\left(
x{\left(\lambda\right)  }\right)
\end{equation}
is a constant scalar (i.e. a 4-scalar, as opposed to a relativistic scalar field) where
\begin{equation} \label{w-squared}
w^{\mu}w_{\mu}=\widetilde{F}^{\mu\nu}\widetilde{F}_{\mu\lambda}\left(
\partial_{\nu}S\right)  \left(  {\partial^{\lambda}}S\right)  =\left(
{{\mathbf{B\cdot}}}\nabla S\right)  ^{2}-\left(  {{\mathbf{E}}\times\nabla
}S{-{\mathbf{B\partial}}}S{{\mathbf{/}}}\partial t\right)  ^{2}\,\,.
\end{equation}
Then it is clear from Eq.~(\ref{u-squared}) that up to an arbitrary universal constant one must set the function $f$ to
\begin{equation} \label{f ito w}
f\left(  x{\left(  \lambda\right) }\right)  =\frac{{\sigma}}%
{\sqrt{\left\vert w^{\mu}w_{\mu}\right\vert }}
\end{equation}
where $\sigma=\pm 1$. Then the norm is
\begin{equation} \label{u^2}
u^{\mu}u_{\mu}={\rm sign}\left(  w^{\mu}w_{\mu}\right) ={\rm sign}\left(  {1-}v^{2}\right)
\end{equation}
which is 1 in the sub-luminal segments of the trajectory, and -1 in the superluminal segments. Eqs.~(\ref{w}) and (\ref{f ito w}) now
give the desired solution for the 4-velocity in terms of the external fields:
\begin{equation} \label{u}
\begin{split}
u^{\mu}   \equiv\left(  \frac{dt}{d\lambda}{,} \frac{d\mathbf{x}}{d\lambda}\right)
    &=\frac{\sigma\widetilde{F}^{\mu\nu }\partial_{\nu}S}{\sqrt{\left\vert \widetilde{F}^{\alpha\beta}\widetilde {F}_{\alpha\gamma}\left(
        \partial_{\beta}S\right)  \left(  \partial^{\gamma}S\right)  \right\vert }}\\
    &=\frac{\sigma\left({\bf B}\cdot\nabla S,{\bf E}\times\nabla
        S - {\bf B}\partial S/ \partial t\right)}{\sqrt{\left\vert \left({\bf B}\cdot\nabla S\right)^{2}
            -\left({\bf E} \times\nabla S -{\bf B}\partial S/\partial t\right)^{2}\right\vert }}\\
    &={\rm sign}\left(  \sigma{{\mathbf{B\cdot}}}\nabla
        S\right) \frac{1}{\sqrt{\left\vert 1-v^{2}\right\vert }}\left(  {1,{\mathbf{v}}}\right)
\end{split}
\end{equation}
where ${\bf v}$ is given by Eq.~(\ref{v ito E and B}). From Eq.~(\ref{u^2}) one has
\begin{equation} \label{u^2 again}
u^{\mu}u_{\mu}\equiv\left(  \frac{dt}{d\lambda}\right)  ^{2}-\left( \frac{d\mathbf{x}}{d\lambda}\right)  ^{2}={\rm sign}\left(
1-v^{2}\right)  \Rightarrow d\lambda=\left\vert dt\right\vert \sqrt{\left\vert 1-v^{2}\right\vert }
\end{equation}
and therefore, following the particular choice Eq.~(\ref{f ito w}) for $f$, $\lambda$ is now a Lorentz invariant (i.e. with respect to
sub-luminal boosts) parameter that in the sub-luminal segments is just the commonly defined proper time $\tau$, but generalised so
that it remains an ordinal parameter throughout the trajectory. The trajectory is now divided up into a sequence of sub-luminal and
superluminal segments, which designation is Lorentz invariant, and for which the norm is $1$ and $-1$ respectively (The segment
boundary points and the sub-luminal and superluminal labels are Lorentz invariant because the three conditions $v<1,$ $v=1,$ and $v>1$
are Lorentz invariants.) Hence, in each segment $u^{\mu}$ is now a true Lorentz 4-vector.
%
\subsection{Limit that the trajectory is a space-filling curve}\label{Space-filling limit}
%
In the limit that the trajectory is sufficiently dense in space, it will no longer be possible to identify individual trajectories,
and one must go over to a continuum description. Then it seems reasonable to assume that at every $x,t$ location (really a small
4-volume centered on that location) observation implies an interaction with the accumulated sum of visits by the trajectory to that
location. In that case one should define a Lorentz vector \textit{field} $j^{\mu}$ say,
\begin{equation}\label{phi}
j^{\mu}\left(x\right) = \int^\infty_{-\infty} d\lambda w^{\mu}=-\int^\infty_{-\infty} d\lambda
\widetilde{F}^{\mu\nu}\partial_{\nu}S\,.
\end{equation}
However, since the trajectory must lie on the surface Eq.~(\ref{E.B=0}) it cannot `fill' 3-space. The best it can do is fill that
surface, in which case $j^{\mu}\left(x\right)$ is defined only on that surface, and is otherwise zero. Since the trajectory is
`conserved' it follows that the 4-divergence of its accumulation vanishes:
\begin{equation}\label{dw=0}
\partial_{\mu}j^{\mu}=-\int^\infty_{-\infty} d\lambda\left(\left(  \partial_{\mu}\widetilde{F}^{\mu\nu}\right)
\left(  \partial_{\nu}S\right)  -\widetilde{F}^{\mu\nu}\partial_{\mu}%
\partial_{\nu}S\right)=0\,\,,
\end{equation}
i.e. $j^{\mu}$ is a 4-current, in agreement with the identification made by Stuckelberg \cite{2:stuckelberg 2}. (That the first term
is zero follows from Maxwell's equation $\partial_{\mu}\widetilde{F}^{\mu\nu}=0$ \cite{32:jackson book}, and the second term is zero
due to the anti-symmetry of $\widetilde{F}$.) Consequently
\begin{equation}\label{j}
j_{0}\left(x\right)=\rho\left(x\right)=\int^\infty_{-\infty} d\lambda \mathbf{B\cdot}\nabla S
\end{equation}
is the density of a conserved charge, for which
\begin{equation}\label{j}
{\bf j}\left(x\right)=\int^\infty_{-\infty} d\lambda \left({\bf E}\times\nabla S-{\bf B}\partial S / \partial t\right)
\end{equation}
is the current density.
%
\subsection{CPT}\label{sec:CPT and other invariants}
%
There is some overlap here with the theory of tachyons, and much of the following may be inferred from the `Reinterpretation
Principle' of tachyon theory \cite{33:recami book, 2:stuckelberg 2, 3:feynman positrons}. However, it is to be stressed that we are
dealing with a single `particle' that is intrinsically massless, whereas the study of tachyons is generally concerned with particles
that have intrinsic `transcendental momentum' - the tachyon counterpart to the intrinsic mass of bradyonic ($v<c$) matter. As a
consequence, traditionally the energy for both bradyons and tachyons becomes singular as light speed is approached from either side.
Light-speed therefore constitutes a barrier, making the labels bradyon and tachyon permanent attributes of these particles - unlike
the massless particle that is the focus of this document.

With reference to Fig.~1 wherein a time reversal occurs at point Q, the segments PQ and QR have different signs for $dt/d\lambda$.
However, which sign is attributed to which segment (the direction of the arrow in the figure) is not decided by Eq.~(\ref{v ito E and
B}). Instead, the sense of the trajectory must be instantiated at some point `by hand'. Noticing that Eqs.~(\ref{f(x,lambda)}) and
(\ref{f ito w}) give
\begin{equation}
{\rm sign}\left(  \sigma{{\mathbf{B\cdot}}}\nabla S\right)
={\rm sign}\left(  \frac{dt}{d\lambda}\right)
\label{sign}
\end{equation}
it is clear that ${\sigma=\pm 1}$ is the degree of freedom that permits one to choose the sign of just one segment, the sign of all
other segments on that trajectory being decided thereafter. If there is only one trajectory then the sign of $\sigma$ is a common
factor for the whole action, so this choice will amount to no more than a convention without any physical consequences unless some
(additional) absolute sense specificity is introduced into the dynamics. But if there are multiple, unconnected, trajectories, then
clearly their relative senses will be important.

%
\begin{figure}
\plotone{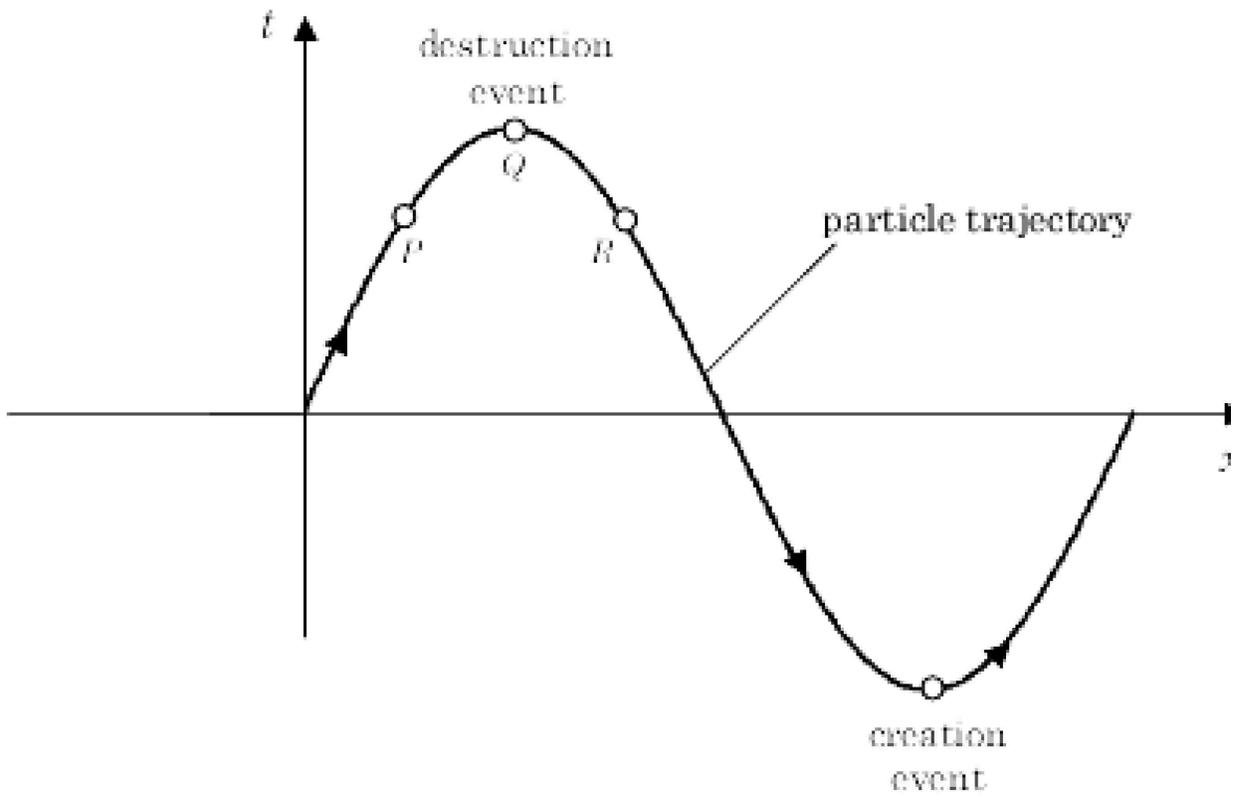}%
 \caption{A trajectory that reverses in ordinary time may be interpreted as giving rise to pair creation and pair
destruction events.}
\end{figure}
%
For given fields at a fixed space-time location $({\bf x},t)$, the change $\sigma\rightarrow-\sigma$ in the equation of motion
Eq.~(\ref{u}) is equivalent to $dt\rightarrow-dt,$ $d{\bf x}\rightarrow-d{\bf x}$. Consistent with CPT invariance therefore, $\sigma$
can be interpreted as the sign of the charge at some fixed point on the trajectory. In accord with the conjecture of Stueckelberg and
Feynman, the alternating segments of positive and negative signs (of $\sigma {{\mathbf{B\cdot}}}\nabla S$) along the trajectory can
then be regarded as denoting electrons and positrons respectively. From the perspective of uniformly increasing laboratory time $t$,
the electrons and positrons are created and destroyed in pairs, as illustrated in Fig.~1. Note that charge is conserved in $t$ just
because these events occur in (oppositely charged) pairs as entry and exit paths to and from the turning points. (If the total
trajectory is not closed, charge is not conserved at the time of the two endpoints of the whole trajectory.)

For the particular case of an anti-clockwise circular trajectory in $x$ and $t$, Fig.~2 identifies the eight different segment-types
corresponding to charge-type, direction in time, direction in space, and speed (sub-luminal versus superluminal). Superluminal, $v>1$,
segments remain superluminal when viewed from any (sub-luminally) boosted frame. Likewise, segments with $v<1$ remain sub-luminal when
viewed from any (sub-luminally) boosted frame. That is, as mentioned above, the labels $v<1$ and $v>1$ are Lorentz invariant. Note
though that the invariant status of these labels is a consequence of the restriction of the boost transformations to sub-luminal
velocities. However, having permitted the massless particle to travel superluminally, one should be prepared to consider augmentation
of the traditional set of transformations to include superluminal boosts of the frame of reference. Upon replacing the traditional
$\gamma$ in the Lorentz transformation formulae with $\gamma=1/\sqrt{\left\vert 1-v^{2}\right\vert }$ and permitting superluminal
boosts (an `extended' Lorentz transformation), the labels $v<1$ and $v>1$ cease to be immutable aspects of the trajectory. The points
$v=1$, however, remain immutable.

The sign of the direction in time of a sub-luminal segment cannot be changed by applying a (sub-luminal) boost transformation and
therefore the sign of the charge is a Lorentz invariant. However, with reference to the labelling exterior to the circle in Fig.~2
(wherein the direction in time is always positive), a superluminally-moving charge \textit{can} change sign under a (sub-luminal)
boost transformation. This is apparent from Fig.~1, where at the pair creation and destruction events $dt/d\lambda=0$, whereas $d{\bf
x}/d\lambda\neq {\bf 0}$ implying that $v=\left\vert d{\bf x/}dt\right\vert $ there is infinite. And if extended Lorentz
transformations are permitted, then \textit{no} part of the trajectory can be given an immutable label corresponding to the sign of
charge.

\begin{figure}
\plotone{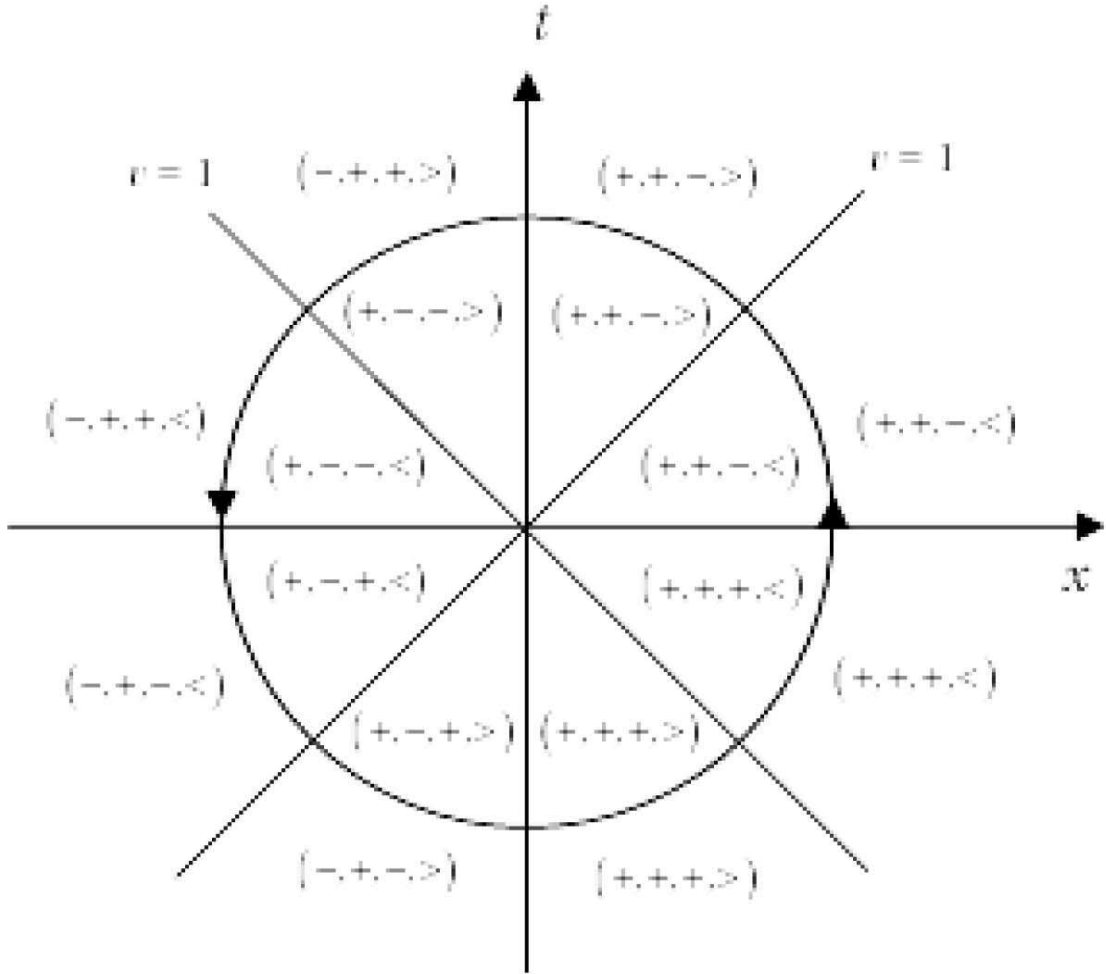}%
\caption{The bracketed symbols denote (sign of charge, sign of $dt/d\tau$, sign of $dx/d\tau$, speed: $>$ or $<$ speed of light). In
the interior of the circle the sign of the charge is fixed, but can take either value in the absence of any other context- the choice
that it is positive is arbitrary. In the exterior of the circle, the bracketed symbols denote the CPT-invariant alternative
designation in which $dt/d\tau$ is always positive.}
\end{figure}
The hypersurface ${\bf E}\cdot{\bf B}=0$ is a Lorentz-invariant collection of events arising here from the requirement that the
determinant of $F$ vanish. One might ask of the other Lorentz invariant associated with the field strengths, $E^{2}-B^{2}$, and why
not the hypersurface $E^{2}-B^{2}=0$ instead? It can be inferred from the fact that the latter quantity is the determinant of
$\widetilde{F}$ that the source of the broken symmetry lies in the fact of the existence of the electric charge but not magnetic
charge; the trajectory of a magnetic charge, were it to exist, would be constrained to lie on the hypersurface $E^{2}-B^{2}=0$.
%
\section{Dynamics}\label{sec:Dynamics}
\subsection{Power flow}\label{sec:Power flow}
%
Whilst following the instructions of the EM field, the particle generates its own advanced and retarded secondary fields as a result
of its motion as determined by the usual EM formulae. By taking the scalar product of
Eq.~(\ref{v ito E and B}) with ${\bf v}$, one observes that%
\begin{equation}
{\mathbf{v}}\left(  x{\left(  \lambda\right)  }\right)  {\mathbf{\cdot E}%
}\left(  x{\left(  \lambda\right)  }\right)  =0 \label{v.E}%
\end{equation}
from which it can be concluded that the massless charge cannot absorb energy from the fields. (Of course, if the system were properly
closed, one could not arbitrarily pre-specify the fields; the incident and secondary fields would have to be self-consistent.) The
massless charge cannot absorb energy from the field because there is no internal degree of freedom wherein such energy could be
`stored'.
%
\subsection{Acceleration}\label{sec:Acceleration}
%
From Eq.~(22) the proper acceleration of the massless charge is
\begin{equation} \label{a}
a^{\mu}=\frac{du^{\mu}}{d\lambda}= u^{\kappa}\partial_{\kappa}u^{\mu}
\end{equation}
where $u$ is given in Eq.~(\ref{u}) and where the factor of $\sigma^{2}=1$ has been omitted. So that the motion is defined, both
$\mathbf{E}$ and $\mathbf{B}$ must be non-zero, or else it must be assumed that one or both must default to some noise value. One
might then ask if the space part of the proper acceleration is correlated with the
Lorentz force, i.e. whether%
\begin{equation}
{{\mathbf{F\cdot a}}}={\left(  {{\mathbf{E}}+{\mathbf{v}}\times{\mathbf{B}}%
}\right)  {\mathbf{\cdot}}}d{{{{\mathbf{u}}}}}/d\lambda\label{F dot a}%
\end{equation}
is non-zero. But it is recalled that the massless particle executes a path upon which the Lorentz force is always zero. Specifically,
from Eq.~(\ref{v ito E and B}),
\begin{equation} \label{v X B}
\begin{aligned}
{\bf v}\times{\bf B}
    &=\frac{\left(  {\bf E}\times\nabla S - {\bf B}\partial S /\partial t\right)  \times {\bf B}} {{{\mathbf{B\cdot}}}\nabla S} \\
    &=\frac{\left(  {{\mathbf{E}}\times\nabla }S\right)  \times{\mathbf{B}}}{{{\mathbf{B\cdot}}}\nabla S} \\
    &= \frac{\left( \mathbf{E}\cdot\mathbf{B}\right)  \nabla S{-}\left(  {{\mathbf{B\cdot}}}\nabla S\right)
        {{\mathbf{E}}}}{{{\mathbf{B\cdot}}}\nabla S} \\
    &=-{\mathbf{E}}
\end{aligned}
\end{equation}
and therefore ${\mathbf{E}}+{\mathbf{v}}\times{\mathbf{B}}={\mathbf{0}}$, and obviously therefore, ${{\mathbf{F\cdot a}}}=0$. It is
concluded that the proper 3-acceleration is always orthogonal to the applied force.
%
\subsection{Motion near a charge with magnetic dipole moment}\label{sec:Motion near a charge with magnetic dipole moment}
%
As an example of a one-body problem, i.e. of a test charge in a given field, we here consider a static classical point charge with
electric field in SI units
\begin{equation} \label{E Coulomb}%
{\mathbf{E}}=\frac{e\hat{\mathbf{r}}}{4\pi \epsilon_0 r^{2}}
\end{equation}
that is coincident with the source of a magnetic dipole field of magnitude $\mu$ oriented in the $z$ direction:%
\begin{equation} \label{B dipole}
{\mathbf{B}}=\frac{\mu_0 \mu\left(  3\hat{\mathbf{r}}z-r\hat{\mathbf{z}}\right) }{4\pi r^{4}}
\end{equation}
(see, for example, \cite{32:jackson book}). Then
\begin{equation} \label{E dot B dipole}%
{\mathbf{E\cdot B}}=\frac{\mu_0 e\mu z}{8\pi^{2}\epsilon_0 r^{6}}\, ,
\end{equation}
and so the constraint that the particle trajectory be confined to the surface ${\mathbf{E\cdot B}}=0$ demands that $z=0$; i.e., the
particle is confined to the equatorial plane for all time. The gradient in the plane is%
\begin{equation} \label{grad E.B}
\left.  {\nabla\left(  {\mathbf{E\cdot B}}\right)  }\right\vert _{z=0}%
=\frac{\mu_0 e \mu\hat{\mathbf{z}}}{8\pi^{2}\epsilon{_0}\rho^{6}} %
\end{equation}
where, ${\rho=\sqrt{x^{2}+y^{2}}}$. With this, and using that at ${z=0}$, ${{\mathbf{B}}=-\mu_0\mu\hat{\mathbf{z}}/4\pi\rho^{3}}$, one
obtains for the denominator in Eq.~(\ref{v ito E and B})%
\begin{equation} \label{B.grad E.B}%
{\mathbf{B\cdot}}\nabla\left(  {{\mathbf{E{\cdot}B}}}\right)  =-\frac{\mu{_0}^2 e\mu ^{2}}{32\pi^{3}\epsilon{_0} \rho^{9}} \,.
\end{equation}
Since ${\mathbf{E\cdot B}}$ is constant in time, the numerator in Eq.~(\ref{v ito E and B}) is just
\begin{equation} \label{E X grad E.B}
{\mathbf{E}}\times\nabla\left(  {\mathbf{E\cdot B}}\right) =\frac{\mu_0 e^{2}\mu}{32\pi^{3}\epsilon{_0}^2\rho^{8}}\hat{\mathbf{\rho}}
\times\hat{\mathbf{z}}=-\frac{\mu_0 e^{2}\mu}{32\pi^{3}\epsilon{_0}^2\rho^{8}} \hat{\mathbf{\phi}} \, ,
\end{equation}
the latter being a cylindrical polar representation of the vector with basis $\left(
\hat{\mathbf{\rho}},\hat{\mathbf{\phi}},\hat{\mathbf{z}}\right)  $. Substitution of Eqs.~(\ref{B.grad E.B}) and (\ref{E X grad E.B})
into Eq.~(\ref{v ito E and B}) gives that the velocity in the cylindrical basis is ${v_{\rho}=v_{z}=0}$ and
\begin{equation}
v_{\phi}\equiv\frac{d\phi}{dt}\rho=\frac{e\rho}{\mu_0 \epsilon_0 \mu}
\end{equation}
which immediately gives that ${z=0}$, ${\rho=\text{constant}}$, and
\begin{equation}
\dot{\phi}=ec^2/\mu \,.
\end{equation}
So it is found that the massless charge is constrained to execute, with radian frequency $ec^2/\mu$, a circular orbit in the
equatorial plane about the axis of the magnetic dipole, as illustrated in Fig.~3.
%
\begin{figure}
\plotone{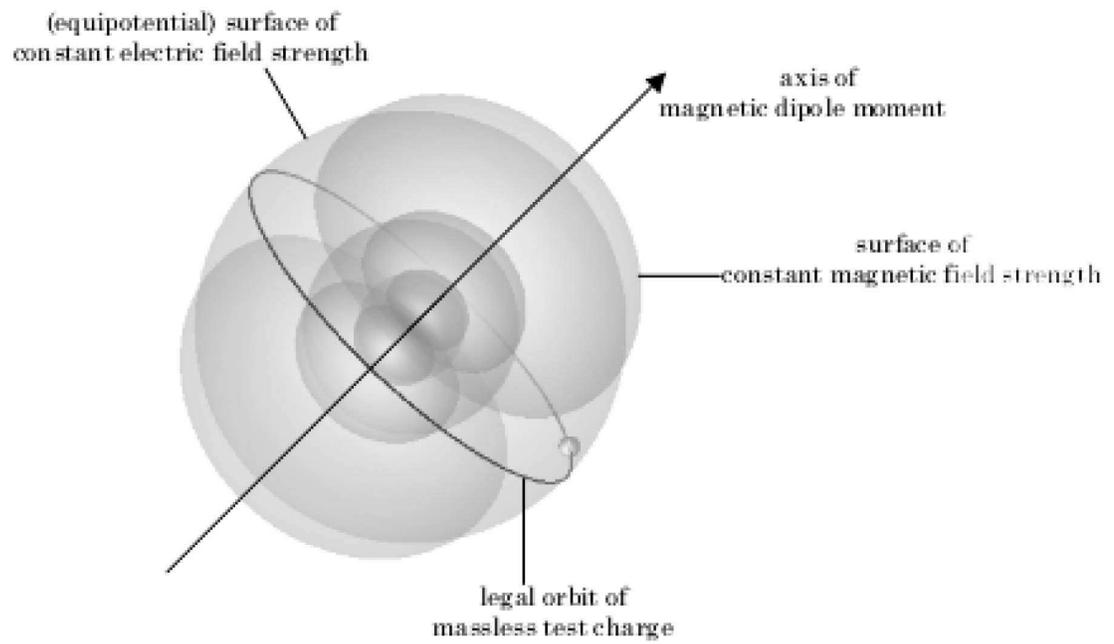}%
\caption{ Orbit of massless charge in a field due to a single electric charge with a magnetic dipole.}
\end{figure}
%

The solution is determined up to two constants: the radius of the orbit, $\rho$, and the initial phase (angle in the $x,y$ plane when
$t=0$). It is interesting to note that if the magnetic moment is that of the electron, i.e. $\mu=ec^2{/2\omega}_{c}$ where
${\omega}_{c}$ is the Compton frequency, then the equatorial orbital frequency is twice the Compton frequency, at all radii.
%
\subsection{Some remarks on the two-body problem}\label{sec:Some remarks on the two-body problem}
%
Previous discussion of the motion of a source has been with the understanding that the fields acting on it are given. Analysis based
upon this assumption may be regarded as the first iteration in an infinite perturbative series, whereupon a completely closed -
non-perturbative - two-body interaction is equivalent to having iterated the particle field interaction to convergence. By two-body
problem, we mean here either a pair of trajectories each of which is forever superluminal or forever sub-luminal, or a single
trajectory with just one light-cone crossing in its entire $\lambda$-history. (A trajectory that crosses its own light cone, emanating
from any 4-point on the trajectory, cannot be regarded as a single charged particle, and should be segmented into multiple particles
accordingly.)

Let the electric field at $\mathbf{r}$ at current time $t$, due to a source at an earlier time $t_{ret}$, i.e., due to a source at
${\mathbf{r}}\left(
{t_{ret}}\right)  $, be denoted by ${\mathbf{E}}_{ret}\equiv{\mathbf{E}%
}\left(  {{\mathbf{r}},t|{\mathbf{r}}\left(  {t_{ret}}\right)  }\right)  $ ,
where $t_{ret}$ is the solution of $t_{ret}=t-\left\vert {{\mathbf{r}}\left(
t\right)  -{\mathbf{r}}\left(  {t_{ret}}\right)  }\right\vert $ . With similar
notation for the magnetic field, the relation between retarded $\mathbf{E}$
and $\mathbf{B}$ fields from a single source can be written ([18])%
\begin{equation}
{\mathbf{B}}_{ret}=\hat{\mathbf{s}}_{ret}\times{\mathbf{E}}_{ret};\qquad
\hat{\mathbf{s}}_{ret}\equiv\frac{{{\mathbf{x}}-{\mathbf{x}}}\left(
t_{ret}\right)  }{| {\bf x}- {\bf x}(t_{ret})|}. \label{B=rXE}
\end{equation}
It is deduced that the retarded fields of a single source give ${\mathbf{B}}_{ret}{\mathbf{\cdot E}}_{ret}=0$ everywhere. In such
circumstances the problem is ill-posed and Eq.~(\ref{v ito E and B}) is insufficient to determine the velocity of a test charge.
Specifically, the component of the velocity of the test charge
in the direction of the $\mathbf{B}$ field is undetermined. In our case however, retarded and advanced fields are mandatory, and the total (time-symmetric, direct action) fields are%
\begin{equation}
{\mathbf{E}}\left(  {{\mathbf{x}},t}\right)  =\frac{1}{2}\left(  {\mathbf{E}%
}_{ret}+{\mathbf{E}}_{adv}\right)  ,\quad{\mathbf{B}}\left(  {{\mathbf{x}}%
,t}\right)  =\frac{1}{2}\left(  {\mathbf{B}}_{ret}+{\mathbf{B}}_{adv}\right) \,\, .
\label{E,B=ret+adv}%
\end{equation}
Their scalar product is
\begin{equation}
\begin{aligned}
{\mathbf{B}}\left(  {{\mathbf{x}},t}\right)  {\mathbf{\cdot E}}\left( {{\mathbf{x}},t}\right)   &  =\frac{1}{2}\left(
{\mathbf{E}}_{ret} +{\mathbf{E}}_{adv}\right)  {\mathbf{\cdot}}\left(  \hat{\mathbf{s}}
_{ret}\times{\mathbf{E}}_{ret}+\hat{\mathbf{s}}_{adv}\times{\mathbf{E}}
_{adv}\right) \label{B.E 2body}\\
&  =\frac{1}{2}\left(  \hat{\mathbf{s}}_{ret}{-}\hat{\mathbf{s}}_{adv}\right) {\mathbf{\cdot}}\left(
{\mathbf{E}}_{ret}\times{\mathbf{E}}_{adv}\right) \nonumber
\end{aligned}
\end{equation}
which is not zero in general, so the massless two-body problem is not ill-posed.

The `no-interaction' theorem of Currie, Jordan, and Sudarshan \cite{34:currie} asserts that the charged particles can move only in
straight lines if energy, momentum and angular momentum are to be conserved; the theorem effectively prohibits {\it any} EM
interaction if a Hamiltonian form of the theory exists. Hill \cite{11:hill}, Kerner \cite{12:kerner}, and others have observed that
the prohibitive implication of the theorem can be circumvented if the canonical Hamiltonian coordinates are not identified with the
physical coordinates of the particles. (Trump and Shieve \cite{10:trump} claim that the original proof is logically circular.) In any
case, it is doubtful that the theorem can be applied to the massless electrodynamics described here: The theorem applies specifically
to direct action classical electrodynamics written in terms of a single time variable, which, if time-reversals are permitted, seems
unlikely to be generally feasible.
%
\section{Discussion and speculation} \label{sec:Discussion}
\subsection{QM-type behaviour}\label{sec:QM-type behavior}
%
It observed that the particle does not respond to force in the traditional sense of Newton's second law. Indeed, its motion is
precisely that which causes it to feel no force, Eq.~(\ref{LF}). Yet its motion is nonetheless uniquely prescribed by the (here
misleadingly termed) `force-fields' $\mathbf{E}$ and $\mathbf{B}$. These fields still decide the particle trajectory (given some
initial condition), just as the Lorentz force determines the motion of a massive particle (again, given some initial condition). But
the important difference is that whereas in traditional (massive) classical electrodynamics the local and instantaneous value of the
external fields determine the acceleration, these fields and their first derivatives determine the \textit{velocity}.

It is also observed that each term in the denominator and numerator of Eq.~(\ref{v ito E and B}) is proportional to the same power
(i.e. cubic) of the components of $\mathbf{E}$ and $\mathbf{B}$. Hence, in the particular case of radiation fields wherein the
magnitudes $E$ and $B$ are equal, the equation of motion of the massless test charge is insensitive to the fall-off of intensity from
the radiating source.

These two qualities of the response to external fields - velocity rather than acceleration, and insensitivity to magnitude - are
shared by the Bohm particle in the de Broglie-Bohm presentation of QM, (\cite{35:de Broglie} - \cite{37:holland}) suggestive, perhaps,
of a relation between the Bohm point and the massless classical charge.

We recall that the Schr\"{o}dinger and (`first quantized') Dirac wavefunctions are not fields in an (a priori) given space-time in the
manner of classical EM, to which all charges respond equally. Rather, the multi-particle Schr\"{o}dinger and Dirac wavefunctions have
as many spatial coordinate triples as there are particles (i.e., they exist in a direct product of 3-spaces). It is interesting that
to some degree this characteristic is already a property of direct action without self action. To see this, note that for two bodies
Eq.~(\ref{lorentz force}) becomes
\begin{equation}
F_{\left(  2\right)  }^{\nu\mu}\left(  x_{\left(  1\right)  }\right)
\frac{dx_{\left(  1\right)  }}{d\lambda}=0,\qquad F_{\left(  1\right)  }%
^{\nu\mu}\left(  x_{\left(  2\right)  }\right)  \frac{dx_{\left(  2\right)  }%
}{d\lambda}=0
\end{equation}
where $F_{\left(  2\right)  }^{\nu\mu}\left(  x_{\left(  1\right)  }\right)  $ is the field at $x_{\left(  1\right)  }\left(
\lambda\right)  $ due to the total future and historical contributions from the particle at $x_{\left( 2\right)  }\left( \kappa\right)
$ such that $\left(  x_{\left(  1\right) }\left(  \lambda\right)  -x_{\left(  2\right)  }\left(  \kappa\right) \right) ^{2}=0$. The
point is that if both particles pass just once through the 4-point $\xi$ say, then, in general, the forces acting on each at that
point are not the same: $F_{\left(  2\right)  }^{\nu\mu}\left(  \xi\right) \neq F_{\left(  1\right)  }^{\nu\mu}\left( \xi\right)  .$
Thus, in common with QM, the fields can no longer be considered as existing in a given space-time to which all charges respond
equally. Instead, each particle sees a different field at the same location. As a result of the conclusion of Section \ref{sec:Self
action}, however, this state of affairs may be subject to revision.

Assuming it has any connection to real physics, the massless particle discussed here cannot be a classical relative of the neutrino.
And it does not seem to be a traditional classical object in need of quantization. Given the fact of the charge, plus the suggestively
QM-type behavior described above, and taking into account the non-locality conferred by super-luminal speeds, it seems more
appropriate to investigate the possibility that the object under consideration is a primitive relative - in a pre-mass and perhaps
pre-classical and pre-quantum-mechanical condition - of the electron.
%
\subsection{Electron mass}\label{ref:Electron mass}
%
One cannot expect convergence with QM or QFT for as long as the intrinsically massless electron has not somehow acquired mass. Though
a detailed explanation will not be attempted here, it is observed that one of the Dirac Large Number coincidences may be interpreted
in a manner suggestive of a role for advanced and retarded fields in the establishing electron mass. It was argued in \cite{30:ibison
ZPF} that the coincidence $m_e \sim e^2/R_H$, where $R_H$ is the Hubble radius, may be a universal self-consistency condition
maintained by EM ZPF fields. Though this may be conceivable in a static universe, it cannot be true in our expanding universe because
the mean path length of a photon of (retarded) radiation is of the order of the Hubble radius. Self-consistency may be possible,
however, if both retarded and advanced fields are employed, as is the case in this document. Then it is conceivable that the
calculations in \cite{30:ibison ZPF} will retain their validity in realistic cosmologies, about which it is hoped to say more
elsewhere. Very briefly, the suggestion is that the electron mass may be the result of constructive interference of time-symmetric
fields reflecting - elastically - off distant sources at zero Kelvin. The idea may be regarded as an extension of the absorber theory
of Wheeler and Feynman, the latter describing the elevated temperature behavior of the same scatterers, which might then serve both as
the explanation for the origin of inertial mass and the the predominance of retarded radiation.
%
\subsection{Self action}\label{sec:Self action}
%
In Section \ref{sec:Action and the Euler equation} the choice was made to deal with infinite self energy by excluding self action by
fiat and adopting the direct action version of EM. The distribution of particle labels in Eqs.~(\ref{total potential}-\ref{total
interaction}) enforces exclusion of the `self-self' terms that connote self-interaction. However, a finding of this investigation is
that a massless particle in a given EM field obeying Eq.~(\ref{u}) can travel at both sub-luminal and superluminal speeds, which
behavior undermines the labelling scheme. To see this, with reference to the left-hand diagram in Fig.~4, if the particle never
achieves light speed, then clearly it will never cross any light cone emanating from any point on that trajectory. That is, the
particle will never see its own light cone. Similarly for a particle that is always superluminal. But with reference to the right-hand
diagram in Fig.~4, a trajectory with both sub-luminal and superluminal segments necessarily intersects its own light cone. If the
whole trajectory is deemed to be non-self-interacting, in accordance with the fiat of no self action, then these points of
electromagnetic contact cannot contribute to the action. Yet these points of interaction are similar in character to the `genuine' -
and therefore admitted - points of contact between any two different trajectories (if indeed there are multiple, distinguishable
trajectories, each with their own starting points and end points). The problem is that the `no self action' rule, necessary for
masslessness of the bare charge, now impacts points of contact that are quite different to the infinitesimally local self action, i.e.
$y=x$ in Eq.~(\ref{single potential}), that was the original target of the rule. In order that these distant points on the same
trajectory conform to the fiat and be excluded from self action it must be supposed that the trajectory, even after any number of time
reversals, forever distinguish itself from other trajectories across all space-time, which requires that each trajectory carry a
unique label (quite apart from its charge and state of motion). In addition to its intrinsic ugliness, this strategy is unappealing
because it precludes the possibility that all electrons and positrons can be described by just one trajectory.
\begin{figure}
\plotone{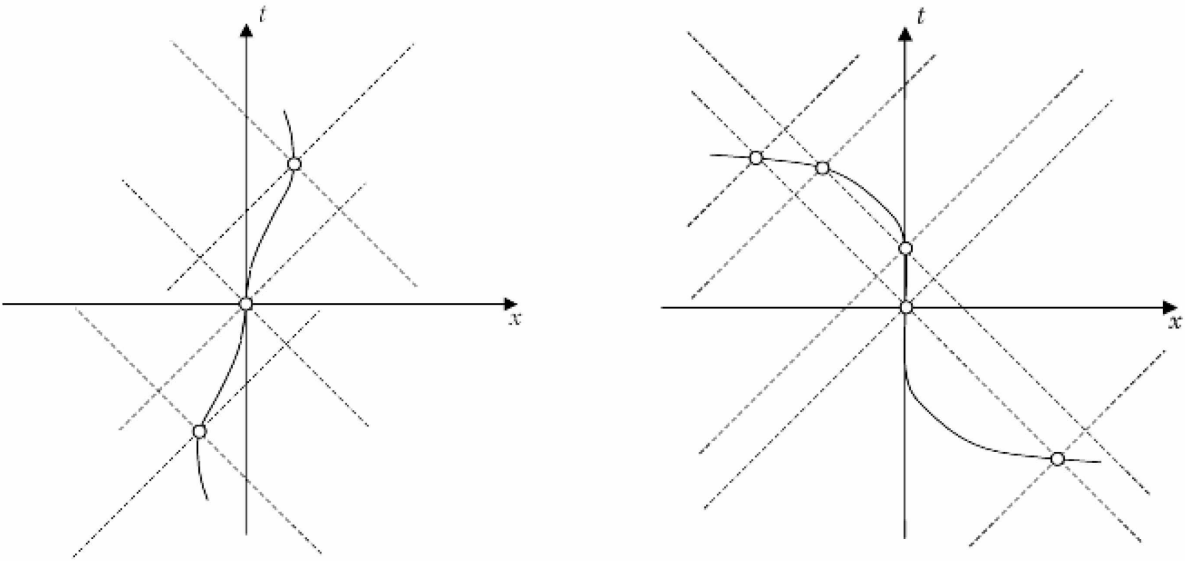}%
\caption{Left-hand diagram: Sub-luminal trajectory showing light cones from three selected space-time points. Right-hand diagram:
Presence of both sub-luminal and superluminal speeds necessarily gives rise to self-interaction, shown here by dashed lines connecting
selected space-time points (circles) on the trajectory.}
\end{figure}

In order to save the masslessness conjecture the alternative must be considered that electromagnetic contact is permitted between
distant points on the \textit{same} trajectory, whilst the energy associated with contact at infinitesimally local points - the
Coulomb self-energy in the rest frame of the particle - is somehow rendered finite. But then it would seem that we are back with the
infinite self-energy problem we set out to avoid by assuming directly-coupled massless sources without self-action, i.e., back to the
problem of finite structures (that are not observed), and Poincar\'{e} stresses (that connote new non-electromagnetic forces) which
have had to be abandoned - for recent examples see \cite{22:schwinger} and \cite{38:ibison mass}. As observed by Pegg \cite{19:pegg},
despite the contention in \cite{8:feynman absorber}, Feynman subsequently decided that it was unlikely that a successful theory could
be constructed without having electrons act upon themselves \cite{39:feynman QED}. In support of this conclusion, Feynman cites the
contribution to a total scattering amplitude from a process in the vicinity of an existing electron in which an electron-positron pair
is created, the latter then annihilating the original electron, leaving only the new electron surviving. In QED, such a process
occurring in the close vicinity of an electron can be regarded as due to self-action. But it cannot be excluded from the physics,
because the same process is necessary for a description of events involving discrete particles initially separated by large distances
in space. In \cite{16:davies direct action 2} Davies comes to a similar conclusion about the inevitability of self-action.

Here, distinct from those works, at least in so far as they address CED, is the novel input of superluminal motion - a consequence of
the presumed masslessness of the bare charge. Superluminal motion permits the possibility of singular self-interaction between
\textit{distant} (i.e. not `infinitesimally-local') points on the same trajectory, leading in turn to the possibility that the Coulomb
energy may be offset by singular self-attraction between different points on the same trajectory. Initial efforts in this direction,
\cite{40:ibison finite action}, though incomplete, are promising. If that approach is successful, an outcome will be that the
distinguishing labels in Eq.~(\ref{total interaction}) will become unnecessary, and it becomes possible to represent the
\textit{total} current by
\begin{equation}
j^{\mu}\left(  y\right)  =\left\vert e\right\vert \int d\lambda u^{\mu}\left( \lambda\right)  \delta^{4}\left(  y-x\left(
\lambda\right)  \right)
\end{equation}
wherein all electrons and positrons are now segments of a single, closed, time-reversing trajectory.
%
\section{Summary}\label{sec:Summary}
%
We have investigated the accommodation of massless charges within the direct-action without self-action form of classical
electromagnetism. The charges move so as to feel no Lorentz force, which, for given external fields, constrains their location at all
times to the surface ${\bf E}\cdot{\bf B}=0$. These two conditions determine the equation of motion for the particle, given by
Eqs.~(\ref{w}) and~(\ref{w defn}). That equation permits sub-liminal and superluminal speeds, time-reversals, and crossing of the
light-speed `barrier', connoting non-locality and pair creation and destruction. An interesting solution has been given for a one-body
problem.

It was argued that for there to be correspondence with the physics of real electrons, perhaps at a pre-quantum-mechanical level,
inertial mass must originate from external electromagnetic interaction. In agreement with Feynman's later comments, it was shown that
the self-action cannot be excluded by fiat after all. It was concluded that for the charge to retain its intrinsic masslessness some
additional remedy is required so as to render the Coulomb part of the self-action finite, in which, perhaps, superluminal motion could
turn out to be an essential novel ingredient.
%
\section*{Acknowledgements}
%
The author is very happy to acknowledge the important role of stimulating and enjoyable discussions with H. E. Puthoff, S. R. Little,
and A. Rueda.
\bigskip
%


\begin{thebibliography}{000}
%
\bibitem{1:stuckelberg 1}
E. C. G. Stueckelberg, Helv. Phys. Acta \textbf{15} (1942) 23.

\bibitem{2:stuckelberg 2}
E. C. G. Stueckelberg, Helv. Phys. Acta \textbf{14} (1941) 321 and 588.

\bibitem{3:feynman positrons}
R. P. Feynman, Phys. Rev. \textbf{76} (1949) 749.

\bibitem{4:schwarzschild}
K. Schwarzschild, Gottinger Nachrichten \textbf{128} (1903) 132.

\bibitem{5:tetrode}
H. Tetrode, Zeits. f. Phys. \textbf{10} (1922) 317.

\bibitem{6:fokker}
A. D. Fokker, Zeits. f. Phys. \textbf{58} (1929) 386.

\bibitem{7:dirac}
P. A. M. Dirac, Proc. R. Soc. Lond. A \textbf{167} (1938) 148.

\bibitem{8:feynman absorber}
J. A. Wheeler and R. P. Feynman, Rev. Mod. Phys. \textbf{17} (1945) 157.

\bibitem{9:feynman direct action}
J. A. Wheeler and R. P. Feynman, Rev. Mod. Phys. \textbf{21} (1949) 425.

\bibitem{13:hoyle 1}
F. Hoyle and J. V. Narlikar, Annals of Physics \textbf{54} (1969) 207.

\bibitem{14:hoyle 2}
F. Hoyle and J. V. Narlikar, Annals of Physics \textbf{62} (1971) 44.

\bibitem{15:davies direct action 1}
P. C. W. Davies, J. Phys. A \textbf{4} (1971) 836.

\bibitem{16:davies direct action 2}
P. C. W. Davies, J. Phys. A \textbf{5} (1972) 1024.

\bibitem{17:davies absorber}
P. C. W. Davies, J. Phys. A \textbf{5} (1972) 1722.

\bibitem{18:davies book}
P. C. W. Davies, \textit{The Physics of Time Asymmetry}, University of California Press, Berkeley (1972).

\bibitem{19:pegg}
D. T. Pegg, Rep. Prog. Phys. \textbf{38} (1975) 1339.

\bibitem{20:feynman mass}
R. P. Feynman, R. B. Leyton, and M. Sands, \textit{The Feynman Lectures on Physics, Volume II}, Addison-Wesley, Reading (1964) chapter
28.

\bibitem{21:poincare}
H. Poincar\'{e}, Comptes Rendue \textbf{140} (1905) 1504.

\bibitem{22:schwinger}
J. Schwinger, Found. Phys. \textbf{13} (1983) 373.

\bibitem{23:haisch 1}
B. Haisch, and A. Rueda, in \textit{Causality and Locality in Modern Physics}, eds. G. Hunter, S. Jeffers and J.-P. Vigier, Kluwer
Academic, Dordrecht (1998) 171.

\bibitem{24:haisch 2}
B. Haisch, A. Rueda, A. and Y. Dobyns, Annalen der Physik \textbf{10} (2001) 393.

\bibitem{25:haisch 3}
B. Haisch, A. Rueda, A. and H. E. Puthoff, Phys. Rev. A \textbf{49} (1994) 678.

\bibitem{26:rueda 1}
A. Rueda and B. Haisch, Found. Phys. \textbf{28} (1998) 1057-1108.

\bibitem{27:rueda 2}
A. Rueda and B. Haisch, in \textit{Causality and Locality in Modern Physics}, eds. G. Hunter, S. Jeffers and J.-P. Vigier, Kluwer
Academic, Dordrecht (1998) 179.

\bibitem{28:rueda 3}
A. Rueda and B. Haisch, Phys. Lett. A \textbf{240} (1998) 115.

\bibitem{29:matthews}
R. Matthews, Science \textbf{263} (1994) 612.

\bibitem{30:ibison ZPF}
M. Ibison, in \textit{Gravitation and Cosmology: From the Hubble Radius to the Planck Scale}, eds. R. L. Amoroso, G. Hunter, M.
Kafatos, and J.-P. Vigier, Kluwer Academic, Dordrecht (2002) 483.

\bibitem{31:leiter}
D. Leiter, in \textit{Foundations of Radiation Theory and Quantum Electrodynamics}, ed. A. O. Barut, Dover, New York (1980) 195.

\bibitem{32:jackson book}
J. D. Jackson, \textit{Classical Electrodynamics}, John Wiley and Sons, New York (1998).

\bibitem{33:recami book}
E. Recami, in \textit{Tachyons, Monopoles, and Related Topics}, ed. E. Recami, North-Holland, Amsterdam (1978) 3.

\bibitem{34:currie}
D. G. Currie, T. F. Jordan, and E. C. G. Sudarshan, Rev. Mod. Phys. \textbf{35} (1963) 350.

\bibitem{11:hill}
R. N. Hill, in \textit{Relativistic Action at a Distance: Classical and Quantum Aspects}, ed. J. Llosa, Springer-Verlag, Berlin (1982)
104.

\bibitem{12:kerner}
E. H. Kerner, in \textit{The Theory of Action-At-A-Distance in Relativistic Particle Dynamics}, ed. E. H. Kerner, Gordon and Breach,
New York (1972) vii.

\bibitem{10:trump}
M. A. Trump and W. C. Schieve, \textit{Classical Relativistic Many-body Dynamics}, Kluwer Academic, Dordrecht (1999) 333.

\bibitem{35:de Broglie}
L. de Broglie, C. R. Acad. Sci. Paris \textbf{183} (1926) 24.

\bibitem{36:bohm}
D. Bohm, Phys. Rev. \textbf{85} (1952) 165.

\bibitem{37:holland}
P. R. Holland, \textit{The Quantum Theory of Motion}, Cambridge University Press, Cambridge (1993).

\bibitem{38:ibison mass}
M. Ibison, in \textit{Causality and Locality in Modern Physics}, eds. G. Hunter, S. Jeffers, and J.-P. Vigier, Kluwer, Dordrecht
(1998) 477.

\bibitem{39:feynman QED}
R. P. Feynman, Phys. Rev. \textbf{76} (1949) 769.

\bibitem{40:ibison finite action}
M. Ibison, in \textit{Has the Last Word Been Said on Classical Electrodynamics?}, eds. A. Chubykalo, V. Onoochin, A. Espinoza, and R.
Smirnov-Rueda, Rinton Press, New Jersey (2004).

\end{thebibliography}
\end{document}